\documentclass[twocolumn,showpacs,amsmath,pre]{revtex4}
\usepackage{graphicx}
\usepackage{amsmath}
\usepackage{amssymb}
\usepackage{natbib}
\usepackage{subfigure}

\begin{document}

\title{Calculations of the Structure of Basin Volumes for Mechanically
Stable Packings}

\author{S. S. Ashwin$^{1,2}$, Jerzy Blawzdziewicz$^{3}$, Corey
S. O'Hern$^{1,2}$, and Mark D. Shattuck$^{4}$}
\affiliation{$^{1}$ Department of Mechanical Engineering and Materials Science,
Yale University, New Haven, CT 06520-8286}
\affiliation{$^{2}$ Department of Physics,
Yale University, New Haven, CT 06520-8120}
\affiliation{$^{3}$ Department of Mechanical Engineering, Texas Tech
University, Lubbock, TX 74909-1021}
\affiliation{$^{4}$ Benjamin Levich Institute and Physics Department, 
The City College of the City University of New York, New York, NY 10031}

\date{\today}
 
\begin{abstract}
There are a finite number of distinct mechanically stable (MS)
packings in model granular systems composed of frictionless spherical
grains.  For typical packing-generation protocols employed in
experimental and numerical studies, the probabilities with which the
MS packings occur are highly nonuniform and depend strongly on
parameters in the protocol.  Despite intense work, it is extremely
difficult to predict {\it a priori} the MS packing probabilities, or
even which MS packings will be the most versus the least probable.  We
describe a novel computational method for calculating the MS packing
probabilities by directly measuring the volume of the MS packing
`basin of attraction', which we define as the collection of initial
points in configuration space at {\it zero packing fraction} that map
to a given MS packing by following a particular dynamics in the
density landscape.  We show that there is a small core region with
volume $V^c_n$ surrounding each MS packing $n$ in configuration space
in which all initial conditions map to a given MS packing. However, we
find that the MS packing probabilities are very weakly correlated with
core volumes.  Instead, MS packing probabilities obtained using initially
dilute configurations are determined by complex geometric features of
the basin of attraction that are distant from the MS packing.
\end{abstract}

\pacs{
63.50.Lm,
83.80.Fg
61.43.-j,
64.70.ps
}
 
\maketitle 

\section{Introduction}
\label{intro}

In contrast to equilibrium, thermal systems, the structural and
mechanical properties of dense granular materials and other athermal
particulate systems depend strongly on the protocol used to create
them.  For example, a number of studies have shown that the packing
fraction of granular assemblies can vary from values associated with
random loose~\cite{onoda} to random close packing~\cite{torquato} as a
function of the vibration amplitude and tapping
history~\cite{nowak,richard}. In addition, the force chain networks
that form, and thus the shear modulus of granular packings depend on
whether they have been generated via shear, isotropic
compression~\cite{majmudar}, or sedimentation via single-particle or
collective deposition~\cite{menon}.

The protocol dependence in dense granular systems arises from the
nonlinear, dissipative, and frictional contact interactions between
grains~\cite{johnson}.  Despite active research in this area, the
distinct contributions from each of these interactions to protocol
dependence has not been determined. In this manuscript, we will
investigate the protocol-dependence of static granular packings by
focusing on a simple system of frictionless spherical particles that
interact via purely repulsive linear spring and velocity-dependent
damping forces.  For a fixed set of boundary conditions, there are
finite number of distinct mechanically stable (MS) packings of
frictionless particles, which grows exponentially with the number of
particles $N$~\cite{xu}.  MS packings exist as discrete points in
configuration space that are characterized by the packing fraction
$\phi_J$ and $N$ particle coordinates ${\vec R}_J = \{ {\vec r}_1,
{\vec r}_2 \ldots {\vec r}_N \}$ and coincide with local minima of the
density landscape~\cite{foot,weber} (or local minima of the potential
energy landscape with zero potential $V=0$). We have shown recently in
both simulations and experiments that the probabilities with which
these distinct MS packings occur are highly nonuniform and depend
on parameters of the packing-generation protocol including the
compression rate, damping coefficient, and initial packing
fraction~\cite{gao1,gao2}.  However, one cannot yet determine {\it a
priori} which MS packings are the most versus the least probable, much
less calculate the packing probabilities as a function of the 
packing-generation protocol.  

\begin{figure}
\includegraphics[width=0.3\textwidth]{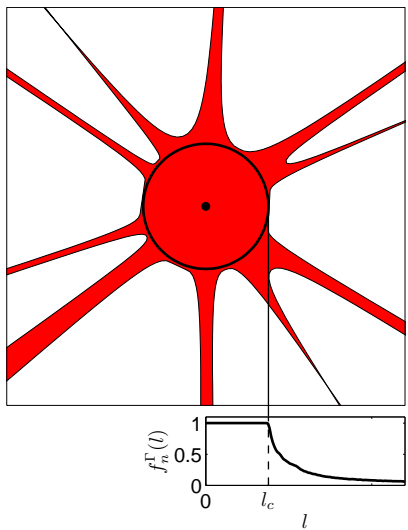}
\vspace{-0.1in}
\caption{(color online) (top) A schematic of the basin of attraction (red) in
$dN$-dimensional configuration space for a typical mechanically stable
(MS) packing (black dot).  (bottom) The corresponding unweighted basin
profile function $f^{\Gamma}_n(l)$ is plotted as a function of
distance $l$ from MS packing $n$ for packing-generation
protocol $\Gamma$.  $f^{\Gamma}_n(l)$ begins to decay from $1$ beyond
an approximately spherically symmetric core size $l_c$, while for $l >
l_c$ the basin is highly branched, thread-like, and $f^{\Gamma}_n(l)
\rightarrow 0$.}
\label{fig1}
\vspace{-0.2in}
\end{figure}

Here we describe a novel computational method for calculating the MS
packing probabilities by directly measuring the volume of the MS
packing `basin of attraction', which we define as the collection of
initial points in configuration space ({\it i.e.}, the red region in
Fig.~\ref{fig1}) at {\it zero packing fraction} that map to a given MS
packing by following a particular dynamics in the density landscape.
Note that our definition of the basin of attraction is
protocol-dependent, and thus the basin volume will vary with the rate
at which energy is dissipated, the compression rate, and other
parameters. In contrast, basins of attraction for glassy
systems~\cite{stillinger,ning2011} are defined as the set of initial
dense liquid configurations that map to the `nearest' local minimum
using steepest descent dynamics at fixed density.  Our definition of
basin volumes is more relevant for granular systems, in which MS
packings are generated from initially dilute configurations.

To aid in the calculation of the basin volumes, we introduce the
unweighted basin profile function $f^{\Gamma}_n(l)$, which is the
fraction of points on a hypersurface in configuration space a distance
$l$ from the $n$th MS packing that maps via a given dynamics (labeled
$\Gamma$) to MS packing $n$. We will show that there is a
hyperspherical core region surrounding each MS packing in which
$f^{\Gamma}_n(l) = 1$ for $l < l_c$, while further from the MS
packing, the basin becomes highly branched, thread-like, and
$f^{\Gamma}(l) \rightarrow 0$.  (See Fig.~\ref{fig1}.) This picture
raises several key questions: 1) Are the MS packing packing
probabilities determined by the size $l_c$ of the core region in
configuration space or dominated by contributions from the thread-like
regions, and 2) do the morphologies of the basins of attraction depend
sensitively on protocol?  We will show below that the MS packing
probabilities are not strongly correlated with the volume of the core
regions in configuration space and are instead determined by features
of the density landscape that are far from each MS packing packing.  Thus,
novel computational geometry techniques~\cite{carlsson} must be developed 
to understand the key features of configuration space that 
control MS packing probabilities.

\begin{figure}
\includegraphics[width=0.35\textwidth]{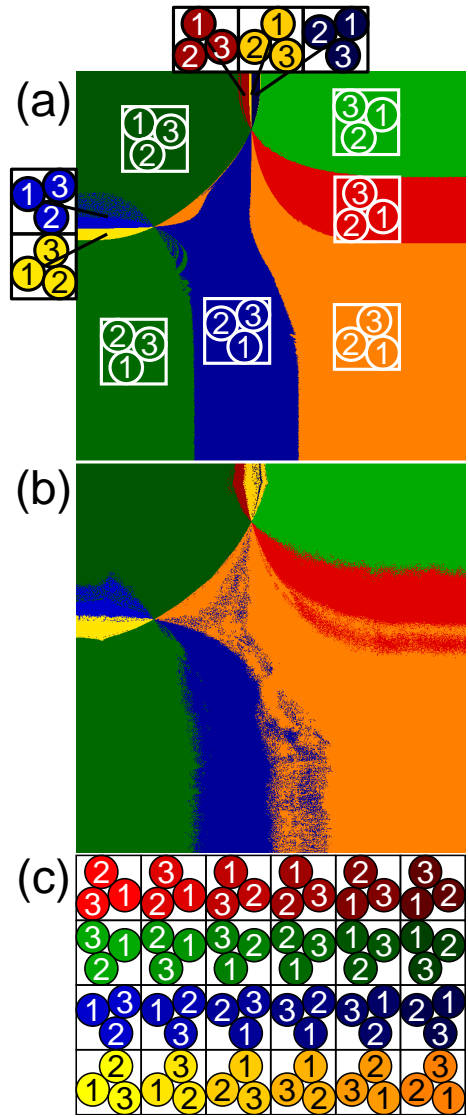}
\caption{(color online) The microstate basins of attraction for a
system of three monodisperse frictionless disks, where particles $2$
and $3$ are initially located at positions $(0.2,0.6)$ and
$(0.45,0.85)$ in the $x$-$y$ plane (with the origin in the lower left
corner). Results are shown for two damping coefficients, (a)
${\widetilde b} =1$ and (b) ${\widetilde b} =0.1$.  The position of
each pixel represents the initial position of particle $1$ and its
color corresponds to one of the $11$ out of $24$ microstates in (c) to
which the system evolved under the compression protocol. For $N=3$
monodisperse systems, there is one distinct MS packing (${\cal N}_s =
1$) with four polarizations (hue; rows) and six permutations
(saturation; columns) for a total of ${\cal N}_m=24$ microstates.  }
\label{fig2}
\end{figure}

\section{Methods}
\label{methods}

To perform our calculations of basin volumes, we focused on a
well-characterized model system composed of $N$ frictionless disks in
2D that interact via purely repulsive linear spring and
velocity-dependent damping forces. $N$ is varied from $3$ to $100$,
and the particles are enclosed in a square cell with fixed walls of
length $L=1$.  Interactions with the walls match those between the
particles. We consider both monodisperse and bidisperse systems, where
the bidisperse mixtures contain half large and half small disks
($N_s=N_l=N/2$) with diameter ratio $\sigma_l/\sigma_s = 1.4$.  

\begin{figure}
\includegraphics*[width=0.7\columnwidth]{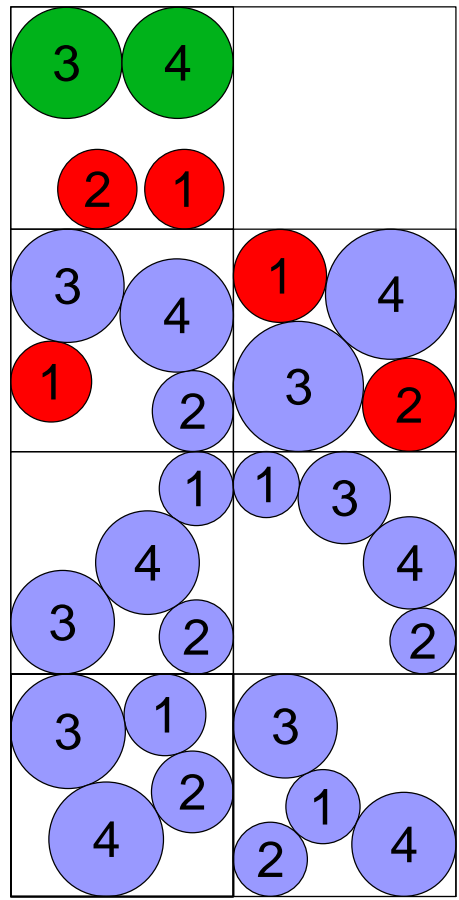}
\caption{(color online) The bottom six configurations are the ${\cal
N}_s=6$ distinct mechanically stable packings for bidisperse systems
with $N=4$. (We will refer to configurations $1$ through $6$ counting
in ascending order from left to right and bottom to top.) The
particles shaded blue form the force-bearing backbone of the
mechanically stable packing.  Particles shaded red are `rattlers'
with fewer than $3$ contacts.  The packing finder generates a small
number of unstable configurations similar to that shown in the upper
left corner with probability less than $0.2\%$, but these are not
included in the analyses.}
\label{4p}
\end{figure}

\begin{figure}
\includegraphics*[width=0.7\columnwidth]{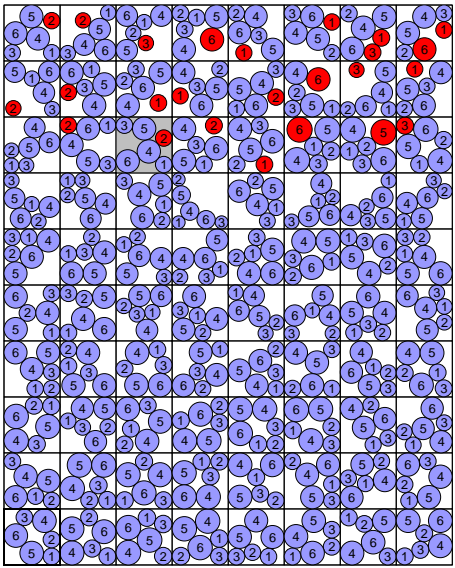}
\caption{(color online) The ${\cal N}_s=80$ distinct mechanically
stable packings for bidisperse systems with $N=6$. The particles
shaded blue form the force-bearing backbone of the mechanically stable
packing.  Particles shaded red are `rattlers' with fewer than $3$
contacts. The unweighted and weighted basin profile functions
are shown in Fig.~\ref{difference} for the configuration in the third 
row that is shaded gray.}
\label{6p}
\end{figure}

In a number of previous studies, we described the MS `packing finder'
that generates a mechanically stable packing via isotropic compression
at $\phi_J$ with infinitesimal overlap from an arbitrary initial
condition at $\phi=0$~\cite{gao1}. Briefly, the algorithm includes the
following steps. For each trial, we initialize the system with random
particle positions inside the unit square at $\phi=0$ and zero
velocities.  We then compress the system in steps of $\Delta \phi =
10^{-4}$ and relax the small particle overlaps after each step by
solving Newton's equations of motion with damping,
\begin{equation}
\label{newton}
m {\vec a}_i = \sum_j {\vec F}(r_{ij}) - b {\vec v}_i,
\end{equation}
where $m$, $\sigma$, and ${\vec a}_i$ are the particle mass, diameter,
and acceleration,
\begin{equation}
\label{force}
{\vec F}(r_{ij}) = \frac{\epsilon}{\sigma} \left(1-\frac{r_{ij}}{\sigma}\right) \Theta\left(1-\frac{r_{ij}}{\sigma}\right) {\hat r}_{ij},
\end{equation}
$\epsilon$ is the characteristic energy of the repulsive spring
interaction, $\Theta(x)$ is the Heaviside step function, ${\widetilde
b} = b \sigma/\sqrt{m \epsilon}$ is the damping coefficient, ${\hat
r}_{ij}$ is the unit vector connecting the centers of particles $i$
and $j$ and $r_{ij}$ is their separation, until the kinetic energy per
particle falls below a specified tolerance $K/\epsilon N < K_{\rm tol}
= 10^{-25}$.  We studied a wide range of values for the damping
coefficient from ${\widetilde b} = 10^{-2}$ to $10$, which mimics
steepest descent dynamics.  The packing-generation algorithm
terminates when the minimized total potential energy per particle
$V/\epsilon N > V_{\rm tol} = 10^{-16}$. As in previous studies on
similar systems with periodic boundary conditions, we distinguish MS
packings based on the spectrum of nontrivial eigenvalues of the
dynamical matrix~\cite{gao2}, and we find that the number of distinct
MS packings ${\cal N}_s$ grows exponentially with $N$ as shown in
Table~\ref{ns}. The ${\cal N}_s=6$ and $80$ distinct MS packings for
$N=4$ and $6$ are shown in Figs.~\ref{4p} and~\ref{6p}.  The packing
finder does produce a small number of unstable packings as shown in
Fig.~\ref{4p}, but these are not included in the analyses.

\begin{table}
\begin{center}
\begin{tabular}{|c|c|c|}
\hline
$N$ & ${\cal N}_s$ & ${\cal N}_m$\\ \hline
$2$ & $1$ & 4\\
$3$ & $1$ & 24\\
$4$ & $6$ & 136\\
$6$ & $80$ & 19440\\
$12$ & $\sim 12000$ & $\sim 4 \times 10^{10}$\\
\hline
\hline
\end{tabular}
\caption{The number of distinct mechanically stable packings ${\cal
N}_s$ and total number of microstates ${\cal N}_m$ versus the number
of particles $N$. For $N=3$ we consider monodisperse systems. For the other 
system sizes, results are given for bidisperse mixtures. For 
$N=12$ we estimate ${\cal N}_s$ and ${\cal N}_m$. We do not include 
unstable packings such as the one in the upper left corner of Fig.~\ref{4p} in 
which the `rigid backbone' of particles can translate.}
\label{ns}
\end{center}
\vspace{-0.25 in}
\end{table}

The fundamental quantity in our approach is the unweighted basin
profile function $f^{\Gamma}_n(l)$ defined as
\begin{equation}
\label{f}
f^{\Gamma}_n(l) = \int d {\vec R} G_{\Gamma}({\vec R},{\vec R}^n_J) \delta(|{\vec R} - {\vec R}^n_J| - l),   
\end{equation}
where $f^{\Gamma}_n(l)$ is sampled on hyperspherical shells a
distance $l$ from MS packing $n$, $\Gamma$ is the specified
compression dynamics, $\delta(x)$ is the Dirac delta function,
$G_{\Gamma}({\vec R},{\vec R}^n_J) = 1$ for points ${\vec R}$ in
configuration space that map to MS packing ${\vec R}^n_J$, and $0$
otherwise. As an illustrative example, we calculate slices of
$G_{\Gamma}({\vec R},{\vec R}^n_J)$ for $N=3$, which has a single MS
packing with ${\cal N}_m$ microstates---$6$ particle-label
permutations and $4$ polarizations obtained by applying all possible
reflections and rotations in 2D consistent with the square cell
boundary conditions~\cite{gao3}.  In Fig.~\ref{fig2}, we plot the
microstate basins of the attraction $\sum_{n=1}^{{\cal N}_m} n
G(\{{\vec r}_1,{\vec r}_2^0,{\vec r}_3^0\},{\vec R}_J^n)$ for fixed
${\vec r}_2^0 = (0.2,0.6)$ and ${\vec r}_3^0 = (0.45,0.85)$.

We calculate the unweighted basin profile function $f^{\Gamma}_n(l)$
using two procedures; the first method is efficient and accurate for
small $l$ and the second for large $l$.  For method $1$, we generate
at least $M=10^6$ points randomly on the surface of a $2N$-dimensional
hypersphere centered on the MS packing with radius $l$.  We then input
each of these configurations as initial configurations into the MS
packing finder with packing fraction $\phi_i=0$.  If a given initial
condition belongs to the basin of attraction of MS packing $n$, the
packing finder will generate packing $n$.  Otherwise, the initial
condition belongs to a different basin.  For the system sizes where we
can achieve complete enumeration, we found that the criterion, $\max_i
(d^j_i - d^k_i)/d^k_i < 10^{-6}$, was sufficiently sensitive to
distinguish MS packings, where $d^j_i$ is the $i$th sorted eigenvalue
of the dynamical matrix for MS packing $j$. From method $1$, the
unweighted basin profile function for MS packing $n$ is
\begin{equation}
f^{\Gamma}_n(l)=\frac{M_n}{M},
\end{equation}
where $M_n$ is the number of initial conditions at $l$ that map to to
packing $n$. 

We define the basin volume for MS packing $n$ generated using 
compression dynamics $\Gamma$ as
\begin{equation}
\label{v}
V_n=\int_{0}^{\sqrt{2N}} S^{\Gamma}_n(l) dl, 
\end{equation}
where 
\begin{equation}
\label{S}
S^{\Gamma}_n(l) = A_{2N}f^{\Gamma}_n(l)l^{2N-1} {\cal P}_n N_s! N_l!
\end{equation}
is the (angle-averaged) weighted basin profile function,
$A_{k}=2\pi^{k/2}/\Gamma(k/2)$ is the surface area of a
$k$-dimensional unit sphere, and ${\cal P}_n$ is the number of
distinct polarizations for MS packing $n$~\cite{gao3}. The probability
of MS packing $n$ for a given compression protocol $\Gamma$ is
proportional to its basin volume, $P^{\Gamma}_n = V^{\Gamma}_n/V_{\rm
tot}$, where $V_{\rm tot} = \sum_{n=1}^{{\cal N}_s} V^{\Gamma}_n = L^{2N}=1$.

Method $1$ becomes extremely inefficient at calculating
$f^{\Gamma}_n(l)$ for large $l>l_c$. Thus, in this regime we implement
method $2$, which was previously employed to calculate the
probabilities $P^{\Gamma}_n$ directly~\cite{xu}. For this method, we
generate at least $10^6$ random points in configuration space and
input these into the packing finder with $\phi_i=0$.  The fraction of
random initial configurations that map to MS packing $n$ determines
$P_n^{\Gamma}$. We can then calculate $f_n^{\Gamma}(l)$ from
$P_n^{\Gamma}$ using Eqs.~\ref{v} and~\ref{S}. Note that an advantage
of method $2$ is that each initial condition provides information
about $P_n^{\Gamma}$ for some $n$ and for $N_s! N_l!$ distances $l$ by
permuting the labels of the final MS packing.

\begin{figure}
\includegraphics*[width=0.85\columnwidth]{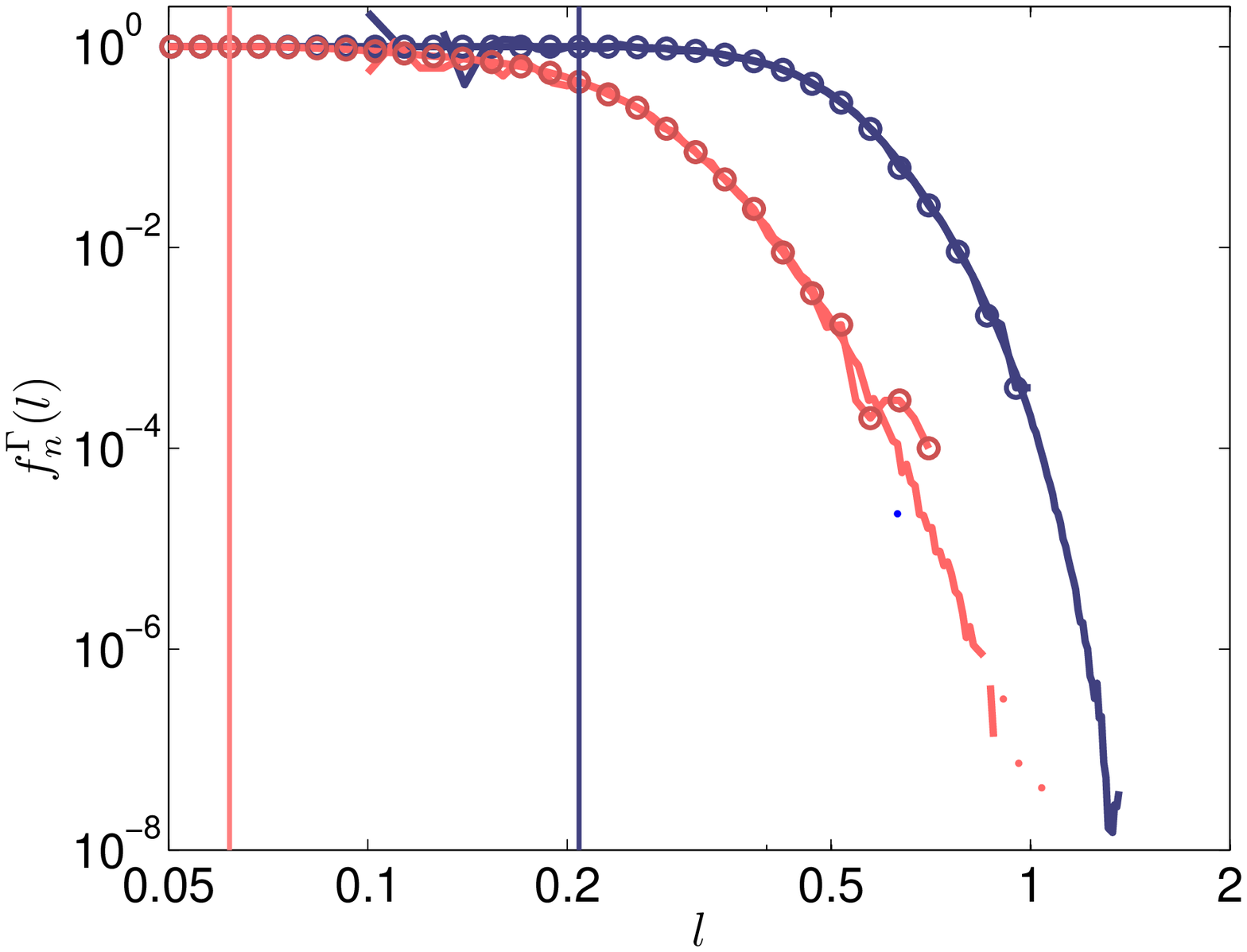}
\includegraphics*[width=0.85\columnwidth]{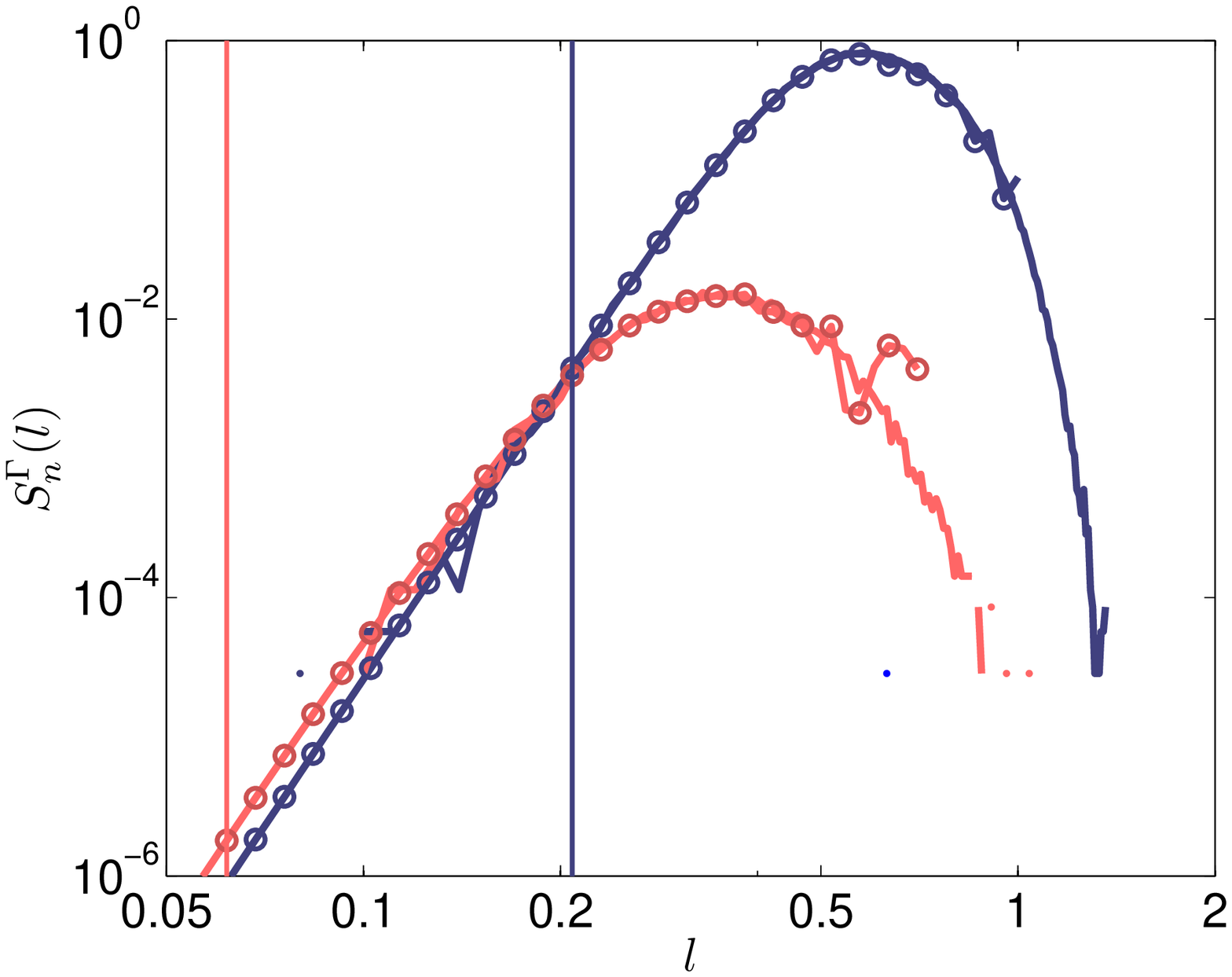}
\caption{(color online) [top] The unweighted $f_n^{\Gamma}(l)$ and
[bottom] weighted $S_n^{\Gamma}(l)$ basin profile functions measured
using methods 1 (circles) and 2 (solid lines) for MS packings $1$
(highest probability; dark blue line) and $4$ (lowest probability;
light red line) shown in Fig.~\ref{4p} plotted on a $\log$-$\log$
scale for $N=4$ and ${\widetilde b}=1$. The vertical lines indicate
$l_c$ for each MS packing.}
\label{figfands}
\end{figure}

\begin{figure}
\includegraphics*[width=0.85\columnwidth]{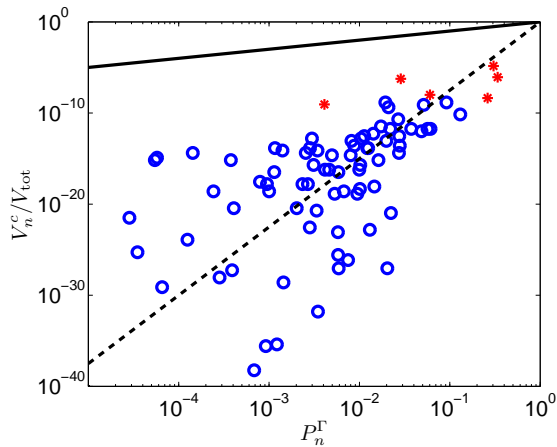}
\caption{The volume $V^c_n$ of the hyperspherical cores surrounding
each MS packing $n$ (relative to $V_{\rm tot}$) plotted as a function
of the MS packing probability $P^{\Gamma}_n$ for each MS packing for
$N=4$ (asterisks) and $6$ (circles) obtained using method $1$ with
damping parameter ${\widetilde b}=1$. The solid (dashed) line has
slope $1$ ($7.5$).}
\label{fig4}
\end{figure}

\section{Results}
\label{results}

Typical basin profile functions $f^{\Gamma}_n(l)$ are shown for the
most and least probable MS packings ($1$ and $4$ in Fig.~\ref{4p}) for
$N=4$ in the top panel of Fig.~\ref{figfands}. For small distances
from the MS packing $l < l_c$, $f_n(l) = 1$.  Beyond the core size
$l_c$, which can vary strongly from one MS packing to another,
$f^{\Gamma}_n(l)$ decays rapidly to zero.  In the bottom panel of
Fig.~\ref{figfands}, we show the weighted basin profile
$S_n^{\Gamma}(l)$ for the same $N=4$ MS packings.  Since
$S_n^{\Gamma}(l)$ is obtained by multiplying $f_n^{\Gamma}(l)$ by
$l^{2N-1}$, the probabilities for obtaining MS packings (when starting
from zero packing fraction) are determined by distances $l > l_c$. For
$N=4$ the average core size is $\langle l_c \rangle \approx 0.1$, the
small particle diameter is $\sigma=0.3$, but the average length scale
that yields $50\%$ of the packing probabilities (near the peak in
$S_n^{\Gamma}(l)$) is $\langle l_p \rangle \approx 0.5$.  We will show
below that $l_p$ grows with increasing system size.  We have validated
the results by ensuring that methods $1$ and $2$ yield the same values
for $f_n^{\Gamma}(l)$ and $S_n^{\Gamma}(l)$ over the range in $l$ in
which the calculations overlap.

In the top panel of Fig.~\ref{figfands}, we show that the core size
for the most probable $N=4$ MS packing is larger than that for the
least probable MS packing, which may suggest that there is a
correlation between the core size and the MS packing probabilities.
To investigate to what extent the hyperspherical core surrounding each
MS packing determines the packing probabilities, we approximate the
basin volume by the volume of a hypersphere of radius $l_c$, $V^c_n =
\pi^Nl_c^{2N}/\Gamma(N+1)$, for each MS packing. In Fig.~\ref{fig4},
we plot $V^c_n/V_{\rm tot}$ versus $P^{\Gamma}_n$ for $N=4$ and $6$.
We find two key results: 1) The volumes $V^c_n/V_{\rm tot}$ are
smaller by many orders of magnitude than the probabilities
$P_n^{\Gamma}$ and 2) A fit to the data for $N=6$ yields $V_n^c/V_{\rm
tot} \sim (P_n^{\Gamma})^{\lambda}$ with $\lambda\approx 7.5$, but
there is only a very weak correlation between $V^c_n/V_{\rm tot}$ and
the packing probabilities. For example, the scatter in the data can
vary by more than $20$ orders of magnitude!  Thus, features of the
basin geometrical structure beyond the core region control the MS
packing probabilities for packings that are generated from dilute
initial configurations.

\begin{figure}
\includegraphics*[width=0.7\columnwidth]{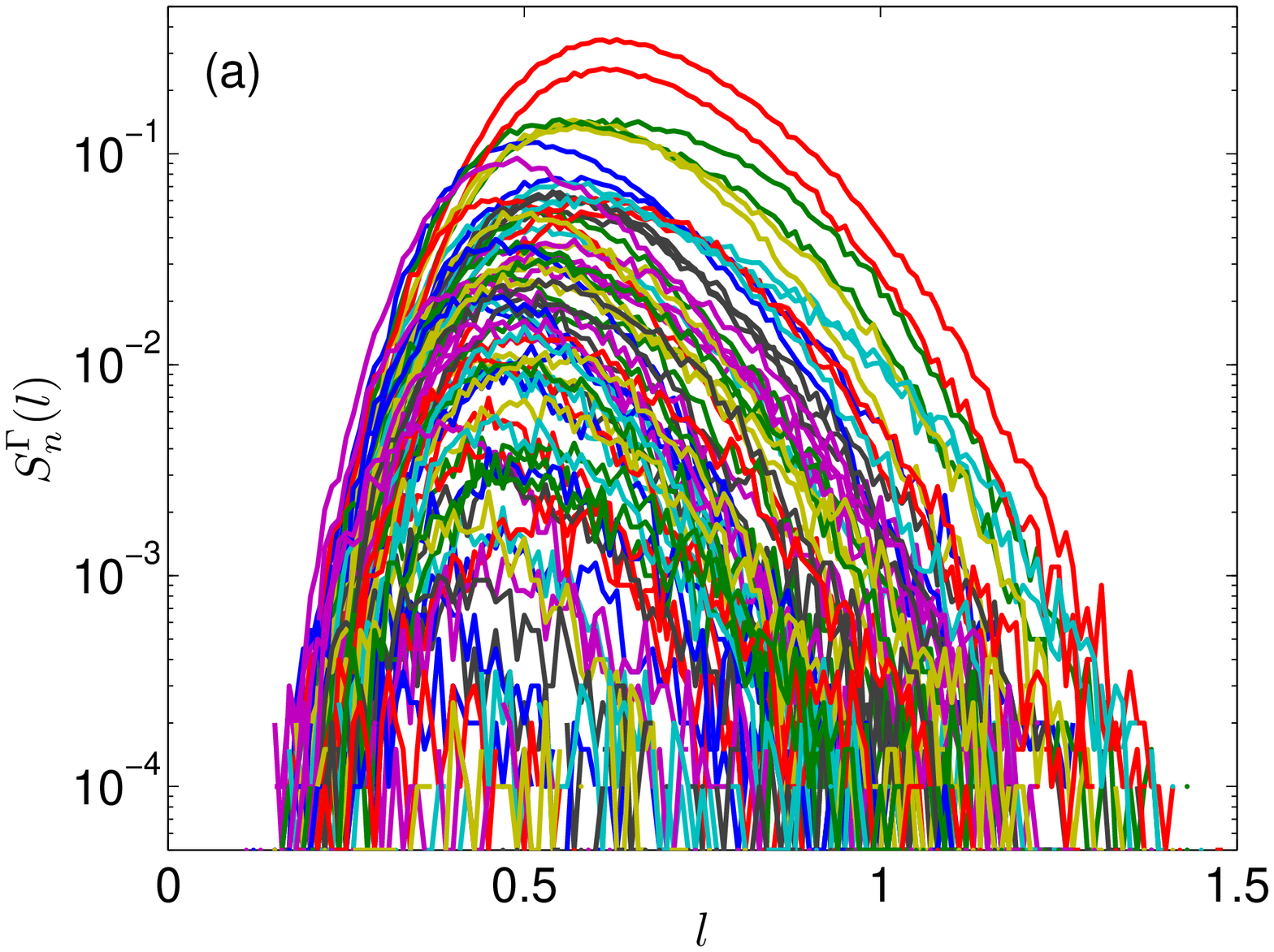}
\includegraphics*[width=0.7\columnwidth]{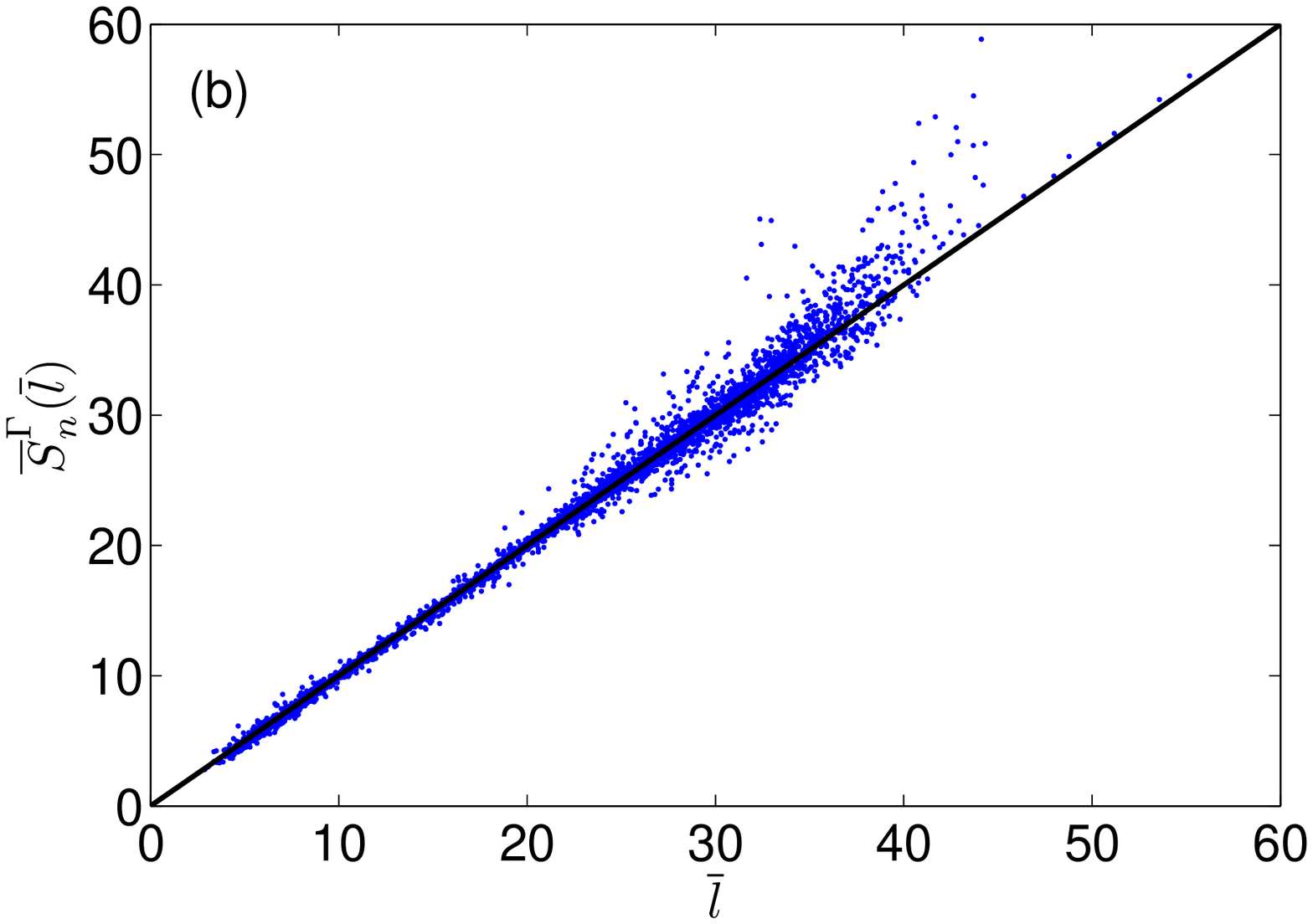}
\caption{(a) The weighted basin profile functions $S_n^{\Gamma}(l)$
(for each of the ${\cal N}_s = 80$ distinct MS packings for $N=6$)
sampled on hyper-spherical shells a distance $l$ from MS packing $n$
using method $2$ with ${\widetilde b}=1$. (b) The scaled weighted
basin profile function ${\overline S}_n^{\Gamma}({\overline l}) =
[S_n^{\Gamma}(l) \theta \Gamma(k) e^{-l/\theta}]^{1/(k-1)}$ plotted
versus the scaled distance ${\overline l}=l/\theta$ for the same data
in (a). The solid line has slope $1$.}
\label{fig3}
\end{figure}

To begin to investigate the nature of the basin morphology beyond the
core region, we characterize in detail the shapes of the weighted
basin profile functions for each of the ${\cal N}_s$ MS packings for
$N=6$ in Fig.~\ref{fig3} (a).  As found for the distribution of
Voronoi volumes in dense granular
packings~\cite{aste,lechenault,kudrolli}, the form of
$S_n^{\Gamma}(l)$ is described by a $\Gamma$-distribution
\begin{equation}
\label{gamma}
S_n^{\Gamma}(l) = \frac{ \left(\frac{l}{\theta}\right)^{k-1} e^{-\frac{l}{\theta}}}{\theta \Gamma(k)},
\end{equation}
where $\theta = (\langle l^2\rangle - \langle l\rangle^2)/\langle
l\rangle$, $k=\langle l \rangle/\theta$, and $\langle l\rangle =
\int_0^{\infty} dl l S_n^{\Gamma}(l)$.  The scaled weighted basin
profile functions ${\overline S}_n^{\Gamma}({\overline l}) =
[S_n^{\Gamma}(l) \theta \Gamma(k) e^{-l/\theta}]^{1/(k-1)}$ for all
microstates collapse when plotted versus the scaled distance
${\overline l}=l/\theta$.  The wider scatter at large ${\overline l}$
is caused by under sampling low probability configurations. We find
similar quality for the collapse at larger $N$.

\begin{figure}
\includegraphics*[width=0.7\columnwidth]{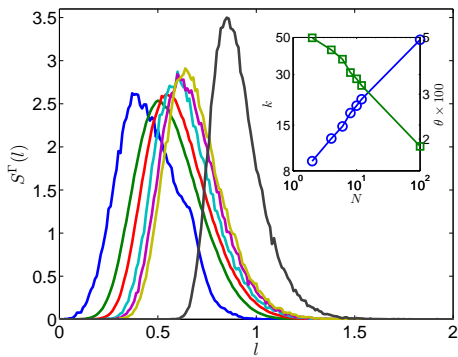}
\caption{The average weighted basin profile function $S^{\Gamma}(l)$
for several system sizes $N=2$, $4$, $6$, $8$, $10$, $12$, and $100$
(from left to right) for damping parameter ${\widetilde b} = 1$. The
inset shows the parameters $k$ (circles; left axis) and $100 \theta$
(squares; right axis) that describe fits of $S^{\Gamma}(l)$ to the
$\Gamma$-distribution (Eq.~\ref{gamma}) versus $N$ on a $\log$-$\log$
scale.}
\label{sn}
\end{figure}

We investigate the system size dependence of the average weighted basin 
profile function 
\begin{equation}
\label{average}
S^{\Gamma}(l) = \sum^{{\cal N}_s}_{n=1} P^{\Gamma}_n S^{\Gamma}_n(l)
\end{equation}
in Fig.~\ref{sn} over the range $N=2$ to $100$. $S^{\Gamma}(l)$ shifts
to larger $l$ with increasing $N$; the peak position $k$ increases by a
factor of $5$ and scales roughly as $\sqrt{N}$ over this range in $N$.
The width $\theta$ slightly narrows over the same range of $N$, scaling roughly
as $N^{-1/4}$.

\begin{figure}
\includegraphics*[width=0.7\columnwidth]{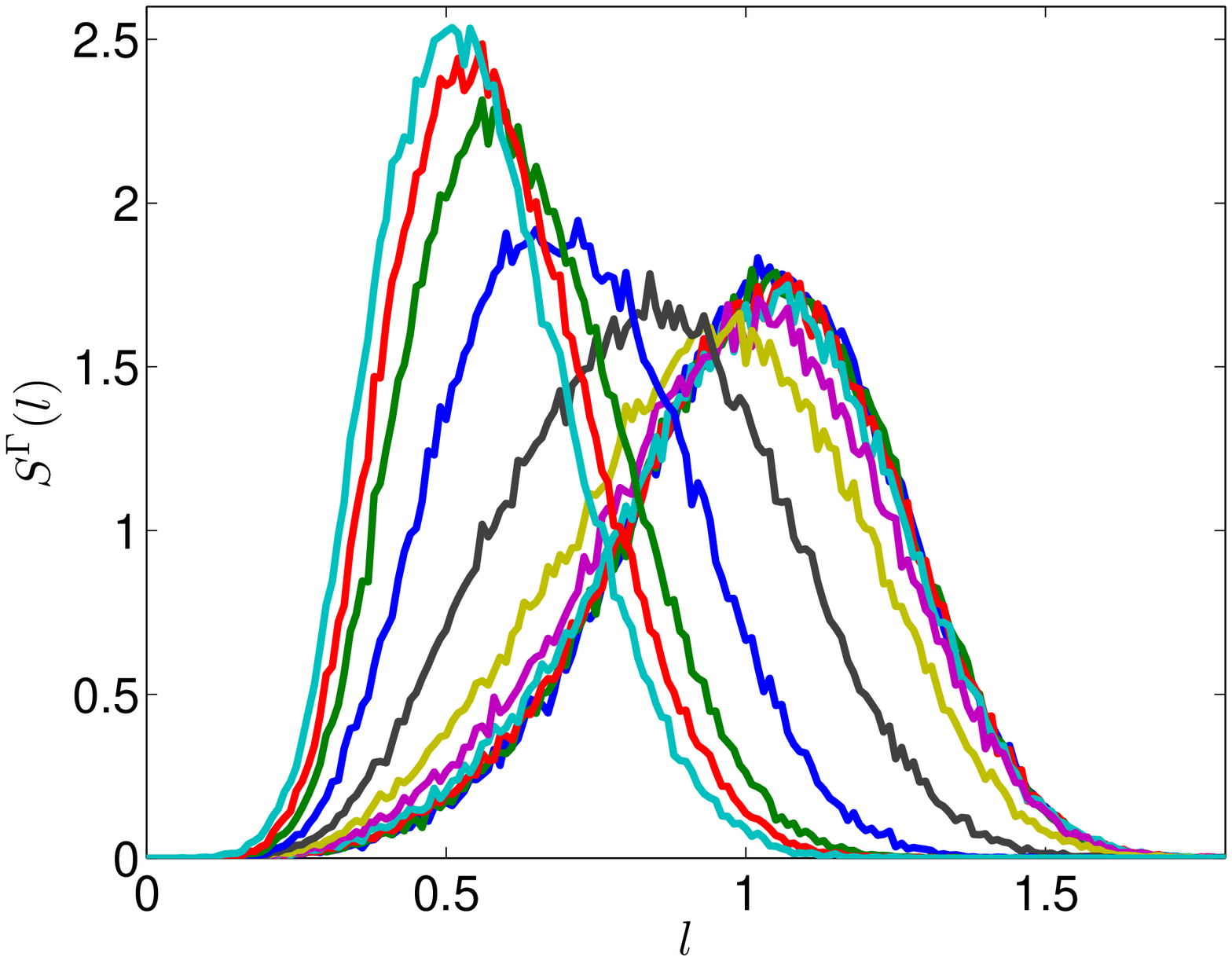}
\includegraphics*[width=0.75\columnwidth]{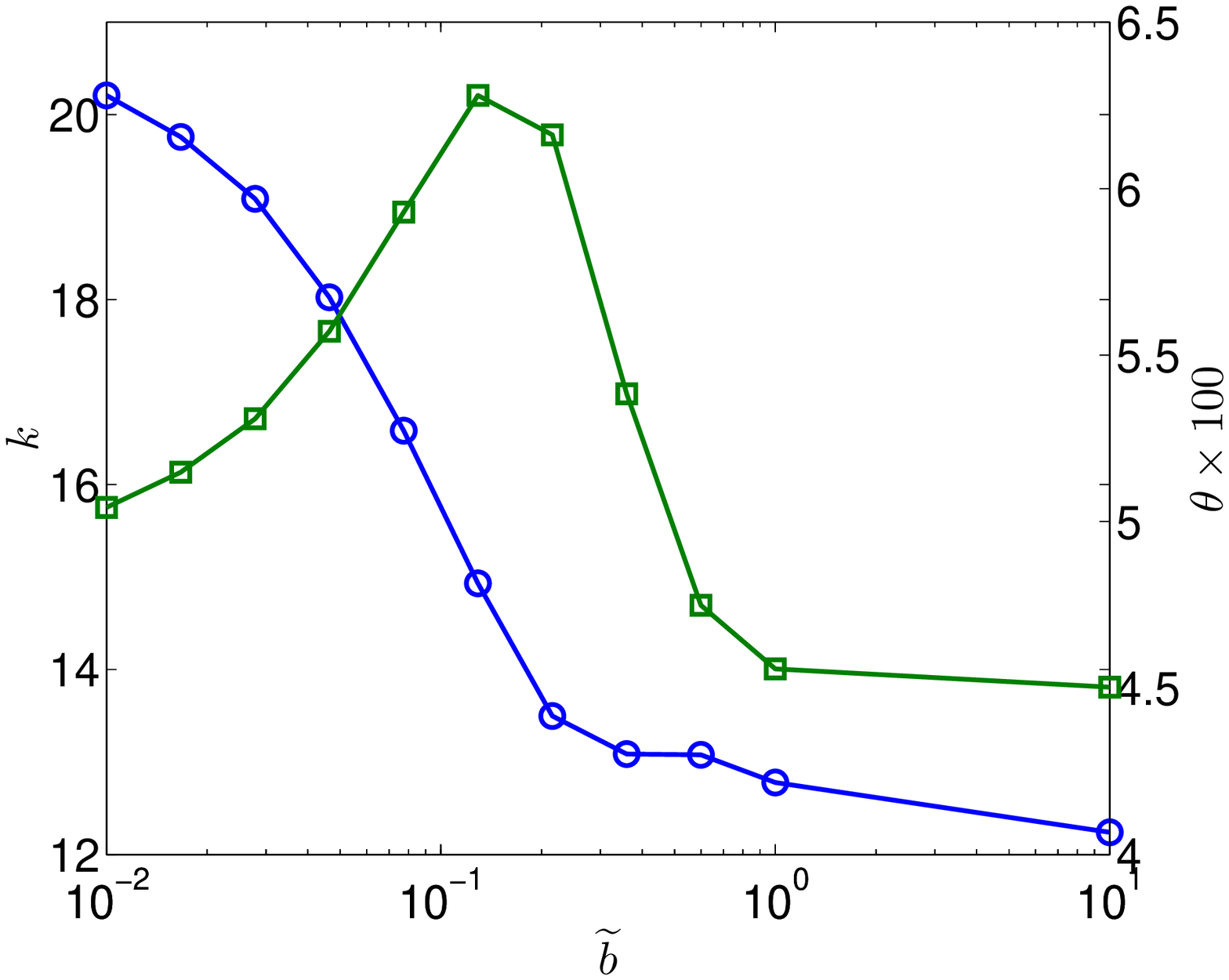}
\caption{[top] The average weighted basin profile function
$S^{\Gamma}(l)$ for $N=4$ plotted over a wide range of the damping
coefficients ${\widetilde b}$ employed in method $2$.  ${\widetilde
b}$ ranges from $0.01$ to $10$ from left to right. [bottom] The
parameters $k$ (circles; left axis) and $100 \theta$ (squares; right
axis) that describe fits of $S^{\Gamma}(l)$ in (a) to the
$\Gamma$-distribution (Eq.~\ref{gamma}) versus ${\widetilde b}$.}
\label{fig5}
\end{figure}

We also investigated the protocol dependence of the basin profile
functions by varying the damping parameter (${\widetilde b}$ in
Eq.~\ref{newton}) used in the packing finder in method $2$.  Lowering
${\widetilde b}$ decreases the rate at which energy is removed from
the system and allows the system to explore larger regions of
configuration space.  In contrast, larger ${\widetilde b}$ increases
the rate at which energy is removed from the system, and thus the
initial configurations are typically closer to the final MS packings.
In the top panel of Fig.~\ref{fig5}, we plot the average weighted
basin profile function versus the damping parameter employed in method
$2$ over three orders of magnitude in ${\widetilde b}$ from $10^{-2}$
to $10$. We were able to saturate the ${\widetilde b}$ dependence of
$S^{\Gamma}(l)$ for both large and small ${\widetilde b}$, {\it i.e.}
for ${\widetilde b} < 10^{-2}$ and ${\widetilde b} > 10$,
$S^{\Gamma}(l)$ is very weakly dependent on ${\widetilde b}$.  The two
parameters $k$ and $\theta$ that describe the shape of $S^{\Gamma}(l)$
exhibit two key features in the bottom panel of Fig.~\ref{fig5}: 1)
the peak of the distribution (captured by $k$) and thus the
lengthscales that determine the MS packing probabilities increase with
decreasing ${\widetilde b}$ and 2) the variance (relative to the
average) depends weakly on ${\widetilde b}$, but does possess a small
peak near ${\widetilde b} = 10^{-1}$.  We expect that the qualitative
features of the ${\widetilde b}$ dependence will persist for larger
system sizes.  In future studies, we will predict the locations of the
peaks in $S^{\Gamma}(l)$ at large and small ${\widetilde b}$ by
calculating the distances between random points in configuration space
and between random points and configurations related by particle label
permutations, respectively.

\section{Conclusions} 
\label{conclusions}

In this manuscript, we described and carried out a novel computational
method for calculating the volume of the MS packing `basins of
attraction', which we define as the collection of initial points in
configuration space at {\it zero packing fraction} that map to a given
MS packing by following a particular dynamics in the density
landscape.  Note that our definition of the basin of attraction is
protocol-dependent, and thus the basin volume will vary with the rate
at which energy is dissipated, the compression rate, and other
parameters.  Using dilute configurations as initial conditions and
including variations in the basin volume with changes in the
packing-generation protocol are crucial for understanding the
protocol-dependent structural and mechanical properties of granular
media and other athermal particulate systems.

\begin{figure}
\includegraphics*[width=0.7\columnwidth]{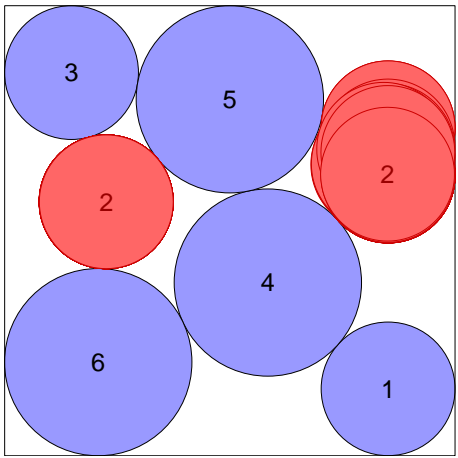}
\caption{$113$ snapshots of one of the $N=6$ mechanically stable
packings (shaded gray in Fig.~\ref{6p}) generated from independent
random initial conditions. This packing contains one rattler particle
(labeled $2$) that can be positioned in the cavity on the left or
right and at multiple positions in the right cavity.}
\label{floater_fig}
\end{figure}

Our computational studies of the basin volumes have uncovered three
important results: 1) A small approximately hyperspherical region of
the basin of attraction with radius $l_c$ surrounds each MS packing,
but the volume of this region (relative to $V_{\rm tot}$) is much
smaller and only very weakly correlated with the MS packing
probabilities in contrast to previous studies of jammed
systems~\cite{ning2011}; 2) the probabilities of MS packings
initialized with dilute configurations are instead controlled by
features of the basins of attraction at lengthscales much further from
the MS packing than the core region.  In addition, the lengthscales
that control the MS packing probabilities grow with increasing system
size and decreasing damping parameter ${\widetilde b}$; and 3) The
shape of the basin profile functions are well characterized by a
$\Gamma$ distribution, which suggests that we can construct a
statistical mechanics-like theory to predict the shape of
$S^{\Gamma}(l)$ and the MS packing probabilities.

\vspace{0.2in}
\noindent {\bf Acknowledgments} This research was supported by the National
Science Foundation under Grant Nos. CBET-0967262 (SA, JB, CO) and
CBET-0968013 (SA, MS).

\appendix 

\section{Rattler Particles}
\label{rattler}

As shown in Figs.~\ref{4p} and~\ref{6p}, MS packings contain rattler
particles.  Two of the ${\cal N}_s = 6$ distinct MS packings for $N=4$
and $24$ of the ${\cal N}_s=80$ distinct MS packings for $N=6$ contain
rattler particles.  For these small-$N$ systems, the fraction of MS
packings that contain rattlers is larger than the fraction of
particles (roughly $5\%$-$10\%$) that are rattlers in large MS
packings~\cite{j}, but these results suggest that the number of 
MS packings containing rattlers is extensive with ${\cal N}_s$~\cite{pair}.
How do rattler particles affect the calculation of the 
basins of attraction for MS packings?  

We find that the correspondence between the unweighted $f_n^{\Gamma}(l)$
and weighted $S_n^{\Gamma}(l)$ basin profile functions breaks down for
small $l$ for MS packings that contain rattler particles.  As shown in
Fig.~\ref{floater_fig}, for MS packings containing rattlers it is
difficult to define uniquely the distance from the initial state to
the final MS packing because the rattler particle can exist over a
range of positions for a given distinct MS packing.  Further, the
different rattler locations may give widely varying contributions to
the MS packing probability.

\begin{figure}
\includegraphics*[width=0.7\columnwidth]{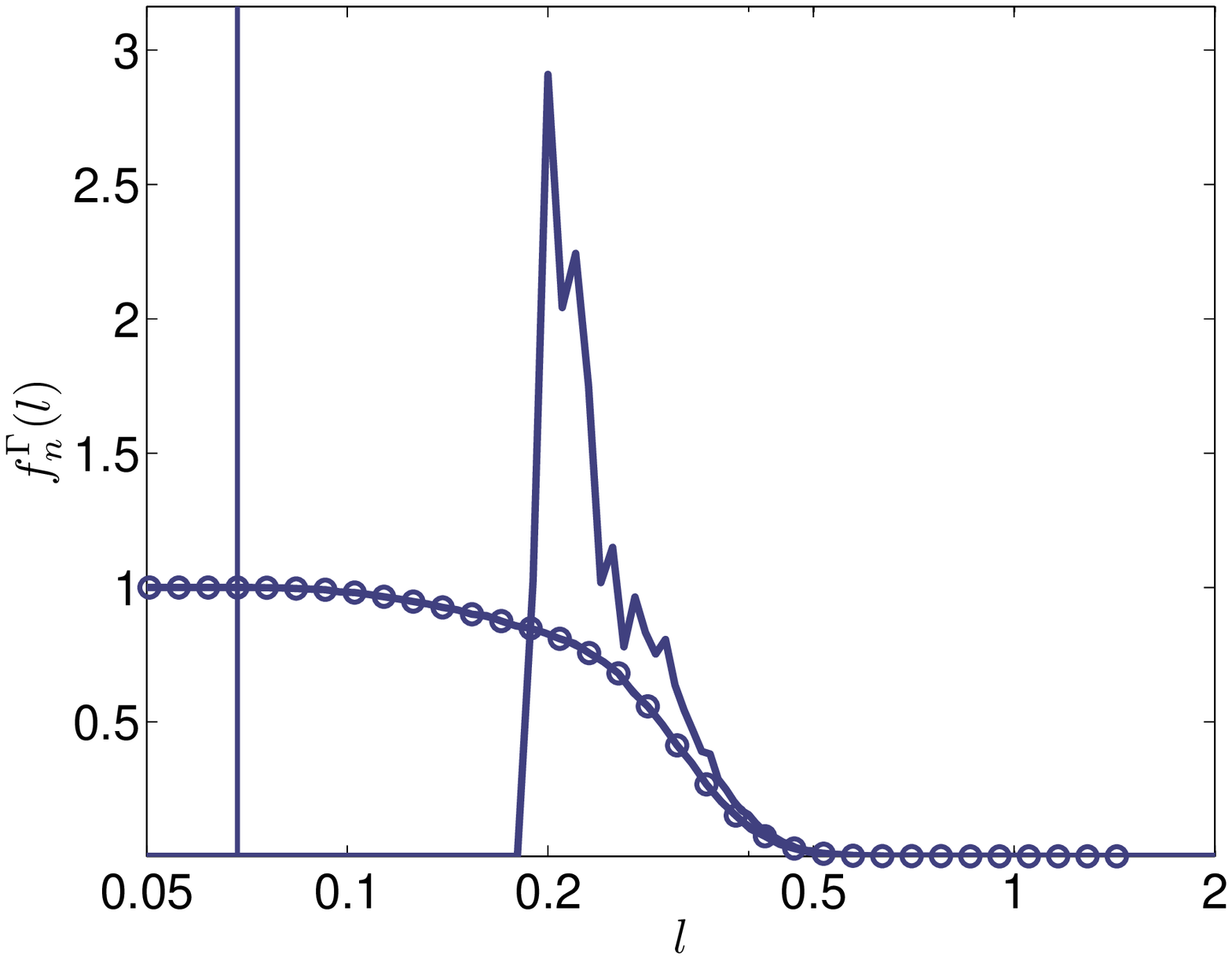}
\includegraphics*[width=0.7\columnwidth]{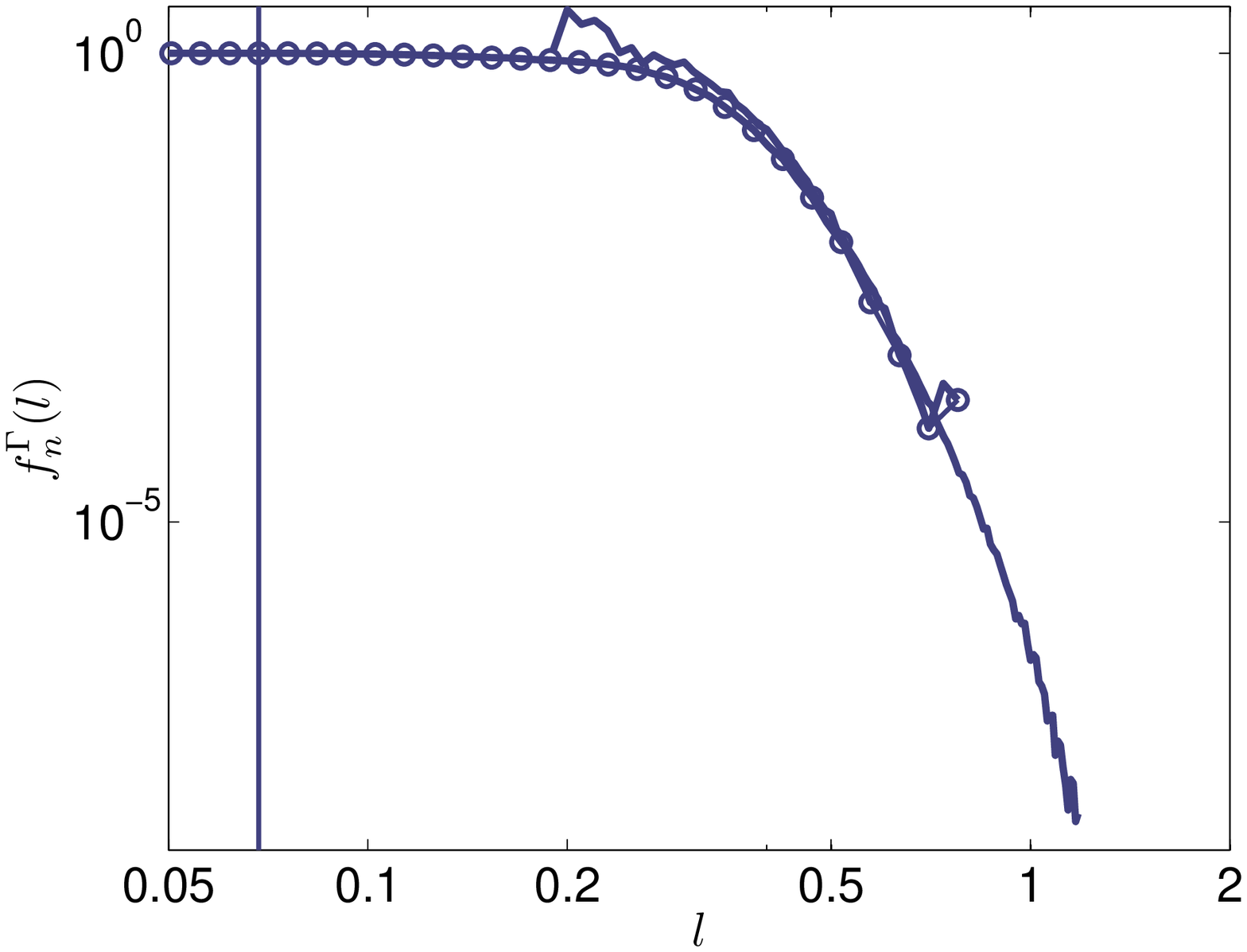}
\caption{(color online) The basin profile function $f_n^{\Gamma}(l)$
on linear-$\log$ [top] and $\log$-$\log$ [bottom] scales measured
using methods 1 (circles) and 2 (solid lines) for the MS packing in
Fig.~\ref{floater_fig} for ${\widetilde b}=1$. The vertical lines
indicate $l_c$.}
\label{difference}
\end{figure}

In Fig.~\ref{difference}, we plot $f_n^{\Gamma}(l)$ calculated using
methods $1$ (circles) and $2$ (solid lines) for the MS packing
depicted in Fig~\ref{floater_fig}.  As described in
Sec.~\ref{results}, for method $1$, we measure the fraction of times
the system returns to the initial MS packing after a perturbation of
size $l$, which is largely unaffected by the presence of rattlers.
For method $2$, we measure the normalized distribution of distances
between the initial configurations and the final MS packings. For
large $l$, method $2$ is also largely unaffected by the presence of
rattlers. However when the initial configuration and final MS packing
are close together ({\it i.e.} small $l$), the fact that the rattler
is not always in the same position in the final MS packing leads to a
significant error in measuring $l$ and hence $f_n^{\Gamma}(l)$, as
shown in Fig~\ref{difference}.  For our measurements of basin volumes
at small $l$, such as $V^c_n$ in Fig.~\ref{fig4}, we show results
using method $1$.  Our main results are insensitive to the presence of
rattler particles because MS packing probabilities (generated from
initially dilute configurations) are determined by features of
$f_n^{\Gamma}(l)$ at large $l$.

\begin{thebibliography}{999}
\bibitem{foot} The density landscape is the surface defined by the
minimum specific volume (or inverse packing fraction) allowed for each
configuration of hard particles.
\end{thebibliography}
\end{document}